\documentclass[12pt,preprint]{aastex}
\usepackage{url}

\shorttitle{Explanation of the sea-serpent magnetic structure in the penumbra}

\title{Explanation of the sea-serpent magnetic structure \\ of sunspot penumbrae}

\author{I. N. Kitiashvili$^{1,2}$, L.R. Bellot Rubio$^3$, A. G. Kosovichev$^{2,4}$, N. N. Mansour$^5$, \\  A. Sainz Dalda$^6$, A. A. Wray$^5$}
\affil{$^1$Center for Turbulence Research, Stanford University, Stanford, CA 94305, USA\\
$^2$NORDITA, Dept. of Astronomy, AlbaNova Univ. Center, SE 10691 Stockholm, Sweden\\
$^3$Instituto de Astrof\'{\i}sica de Andaluc\'{\i}a (CSIC), Apdo. de Correos 3004, E-18080 Granada, Spain\\
$^4$Hansen Experimental Physics Laboratory, Stanford University, Stanford, CA 94305, USA \\
$^5$NASA Ames Research Center, Moffett Field, Mountain View, CA 94040, USA\\
$^6$Stanford-Lockheed Institute for Space Research, Stanford, CA 94305, USA}
\altaffiltext{1}{e-mail: irinasun@stanford.edu}

\begin{document}

\begin{abstract}
Recent spectro-polarimetric observations of a sunspot showed the
formation of bipolar magnetic patches in the mid penumbra and their
propagation toward the outer penumbral boundary. The observations were
interpreted as being caused by sea-serpent magnetic fields near the
solar surface \citep{dalda08}. In this Letter, we develop a 3D
radiative MHD numerical model to explain the sea-serpent structure and
the wave-like behavior of the penumbral magnetic field lines. The
simulations reproduce the observed behavior, suggesting that
the sea-serpent phenomenon is a consequence of magnetoconvection in a
strongly inclined magnetic field. It involves several physical
processes: filamentary structurization, high-speed overturning
convective motions in strong, almost horizontal magnetic fields with
partially frozen field lines, and traveling convective waves. The
results demonstrate a correlation of the bipolar magnetic patches with
high-speed Evershed downflows in the penumbra. This is the first time
that a 3D numerical model of the penumbra results in downward directed
magnetic fields, an essential ingredient of sunspot penumbrae that
has eluded explanation until now.

\end{abstract}

\section{Introduction}
It is well-known that the sunspot penumbra (the outer part of
sunspots) has a very complicate filamentary structure and a strong
non-stationary outflow. This outflow is responsible for the so-called
Evershed effect, a Doppler shift of the spectral lines emerging from
sunspots \citep{evershed1909}. The magnetic structure of the penumbra
can be represented as a mixture of two magnetic field components with
different inclinations and strengths
\citep[e.g.,][]{degenhardt91,schmidt92,title93,lites93,
stanchfield97,bellot04,sanchez05,borrero05,beck08}. However, the radial Evershed
flow (in particular, the high-speed 'Evershed coluds') is associated with
the more strongly inclined, almost horizontal
field \citep[e.g.][]{title93,shine1994,bellot03}.

Recently, significant progress in our understanding of the Evershed
effect was made by numerical simulations \citep{heinemann2007,
scharmer2008,rempel2009,rempel_sci09,kiti09a}. These studies
suggested that the Evershed flow is a consequence of overturning
magnetoconvection in the presence of inclined magnetic fields, and
that the driving mechanism is associated with traveling convective
waves that propagate in the direction of the magnetic field
inclination \citep{hurlburt2000,kiti09a}. The issue is not resolved
and other interpretations are possible
\citep[see][]{schliche09}. Thus, it is important to confront the
simulations with as many observations as possible.

In this Letter, we examine the idea that the sea-serpent penumbral
field lines detected by \cite{dalda08} in high-resolution Hinode
measurements are related to the same mechanism of overturning
convection and traveling convective waves in a strong inclined
magnetic field. Our analysis is based on the radiative MHD simulations
of \citet{kiti09a}.

\section{Moving bipolar magnetic patches in the penumbra and
the sea-serpent model}

Analyses of visible and near-infrared spectro-polarimetric
observations from the ground revealed that in the mid and outer
penumbra the Evershed flows are directed downward and have magnetic
polarities opposite to that of the spot \citep[][see Be\-llot Rubio 2009
for a detailed description]{westend97}. The spectro-polarimeter aboard
Hinode \citep{kosugi07,tsuneta2008} made it possible to identify
isolated magnetic patches of opposite polarity associated with strong,
even supersonic, Evershed downflows
\citep{ichimoto07, bellot09, sanchez09}. This provided a beautiful
confirmation of the results obtained from ground-based observations at
lower resolution. \cite{dalda08} then discovered that the opposite
polarity patches move radially outward, often accompanied by another
patch having the polarity of the spot and being located further away
from the umbra. \cite{dalda05} and \cite{ravindra06} also detected the
evolution of reversed-polarity patches in the penumbra using SOHO/MDI
observations \citep{scherrer1995}, but their bipolar nature could not be
established. Taken together, these results provide an indication that
the moving magnetic features observed in the penumbra and the Evershed
flow belong to the same physical mechanism.

The magnetic patches move between the more vertical filaments of the
penumbra with a typical velocity of $0.3-1$ km/s, their length is 2--3
arcsec, the mean width is $\sim 1.5$ arcsec, and they have a lifetime
of 0.5 -- 7 hours \citep{dalda08}. Sometimes, the bipolar pairs appear
one behind another along the same filament. This can be the signature
of a wave-like behavior, similar to the wave behavior of
magnetoconvection in the penumbra found in numerical simulations
\citep{kiti09a,kiti09b}.

Based on the observational data, \citet{dalda08} proposed a
sea-serpent model which explains the existence of moving bipolar
structures as wavering magnetic field lines. In this model, almost
horizontal field lines return to the solar interior and then come back
to the photosphere, thus forming bipolar patches. This scenario is
similar to that discussed by Schlichenmaier (2002) in the context of
the moving tube model. It is also similar to the model proposed by
\cite{harvey73} for moving magnetic features in the sunspot moat
\citep[MMFs;][]{sheeley1969}. Indeed, it has been suggested that
MMFs represent the continuation of similar features in the penumbra
\citep{dalda05,ravindra06,dalda08,kubo08}.

\section{Simulations of magnetic structure and dynamics of sunspot penumbra}

For numerical simulations of sunspot penumbra conditions we use the 3D
radiative MHD code "SolarBox" \citep{jacoutot08a,jacoutot08b}. This
code takes into account all essential physics and includes sub-grid
scale turbulence modeling based on the large-eddy simulation (LES)
approach. A dynamic Smagorinsky turbulence model provides the best
agreement with observations in terms of the acoustic oscillation
power. However, in these simulations we use a computationally more
efficient hyperviscosity model, because it shows results qualitatively
very similar to the dynamic model. The code uses a real-gas equation
of state, takes into account ionization and excitation of all abundant
species in the LTE approximation, and includes radiative transfer and
magnetic effects \citep{ripoli03,jacoutot08a,kiti09a,kiti09c}.

We simulate the behavior of solar magnetoconvection for an initial
background magnetic field of various strengths, from 600 to 2000~G,
but with the same inclination of $85^\circ$ to the vertical axis. Our
previous results have shown that these condition provide a very good
model for the Evershed effect \citep{kiti09a}. The computational
domain is $6.4\times6.4\times6$~Mm$^3$ and the grid resolution is
50~km. The results were verified using a higher resolution of 25~km
and a larger horizontal size of 25~Mm. The lateral boundary conditions
are periodic, and we keep the mean initial inclination of the field in
the whole box domain by setting up top and bottom boundary
conditions that conserve the total magnetic flux, but locally
the field strength and inclination can freely change.

The magnetic field strongly influences convective motions.  For
instance, in the case of vertical magnetic fields, convective granules
become smaller with increasing field strength
\citep{stein2001} and also move faster, generating more high-frequency
turbulence \citep{jacoutot08b}. A strong inclined magnetic field deforms the
granules, making them elongated along the field lines. When the
field becomes almost horizontal the convective cells form a
filamentary structure, and high-speed flows (reaching 5-6 km/s)
appear in the direction of the magnetic field inclination
\citep{kiti09a}. These flows share many similarities with the
observed Evershed flows, including their nearly-supersonic speeds.

\begin{figure}
\begin{center}
\includegraphics[scale=0.8]{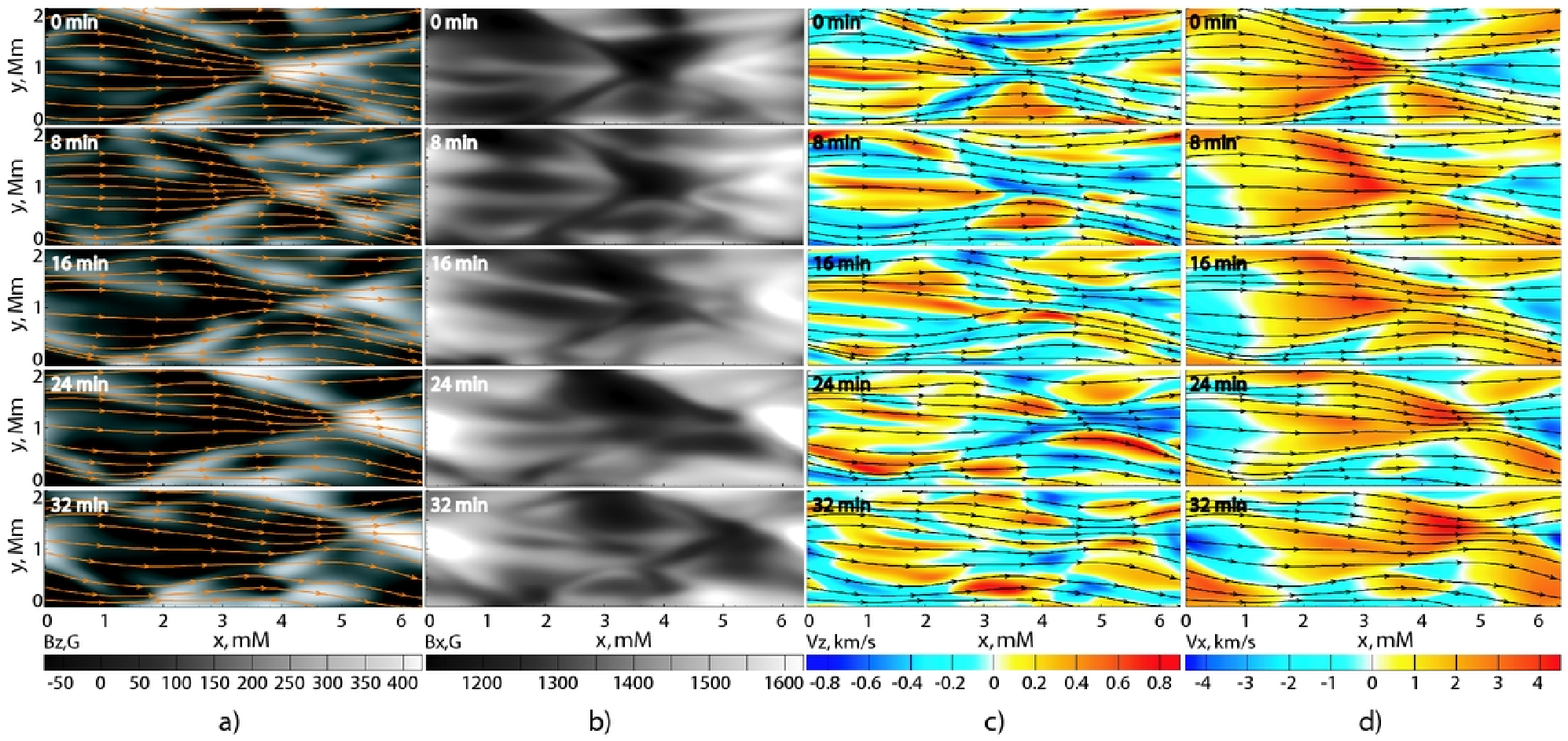}
\end{center}
\caption{Snapshots of the vertical (a) and horizontal (b) components of magnetic field, and the vertical (c) and horizontal (d) components of velocity at the photospheric level for different moments with a cadence of 8 min for the simulations for the initial $B_0=1200$G background magnetic field with the $85^\circ$ inclination angle from the vertical in the $x$-direction. The overlayed curves are projections of magnetic field lines.
\label{2D}}
\end{figure}

For illustration of our results we use a simulation run with initial
magnetic field of $B_0=1200$~G and inclination angle of $85^{\circ}$
pointing in the direction of the horizontal $x$-axis.  Figure~\ref{2D}
shows a time sequence of 2D snapshots of the horizontal and vertical
components of the magnetic field, $B_x, B_z$, and velocity, $v_x,
v_z$, with 8-min cadence for five time moments: 0, 8, 16, 24 and 32
mins. The overplotted curves are projections of the magnetic field
lines.  The black and white areas seen in the vertical magnetic field
(Fig.~\ref{2D}a) correspond to patches of negative and positive
polarities. The horizontal magnetic field component (Fig.~\ref{2D}b)
shows variations of the magnetic filamentary structure, which is a
result of magnetoconvection in the inclined field. The dark areas
correspond to ''magnetic gaps", areas with a weaker (but not zero)
magnetic field, that correspond to observations \citep{severnyi65,
ichimoto07}. It is clear that the areas of weaker horizontal magnetic
field correspond to the transitions between negative and positive
polarities of the vertical field. In these transition regions, strong
downward flows are observed (Fig.~\ref{2D}c). The transition regions
are also characterized by strong $\sim 4-6$ km/s horizontal velocities
(Fig.~\ref{2D}d). The strongest horizontal flow in the direction of
the magnetic field inclination is associated with the negative
polarities. The correlation between downflows and negative polarity
patches is in agreement with the observations, which show that the
downward penumbral motions occur along magnetic fields returning back
to the solar surface \citep{westend97,ichimoto07,bellot09,sanchez09}.

\begin{figure}
\begin{center}
\includegraphics[scale=0.5]{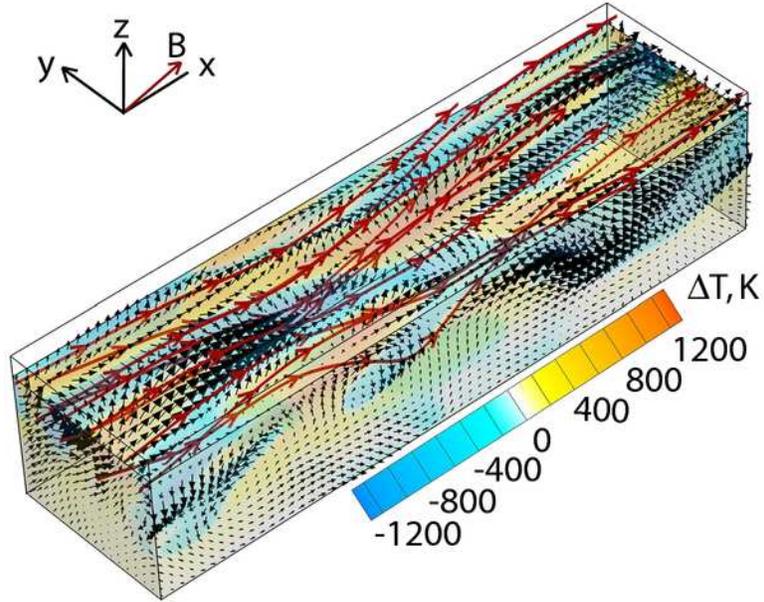}
\end{center}
\caption{3D rendering of temperature variations, magnetic field lines (red curves) and flow field (arrows). \label{3D}}
\end{figure}

Figure~\ref{3D} displays the 3D structure of magnetic field lines (red
curves), temperature variations, and the velocity field (arrows). The
top horizontal slice corresponds to the solar surface. The results
demonstrate a strong coupling between the fluid flow and magnetic
field. In particular, the upflows correspond to magnetic field lines
rising above the surface (positive polarity) while the downward flows
correlate with the negative polarity. This corresponds very well to
the observed properties of the penumbra and the sea-serpent model.

\section{Discussion and Conclusion}

In this Letter, we have used the results of numerical simulations of
magnetoconvection in strong inclined magnetic field to interpret
polarimetric observations of a sunspot penumbra. The results reproduce
the moving bipolar magnetic elements observed in high-resolution
SOHO/MDI and Hinode/SOT data and also their properties, supporting the
sea-serpent model proposed by \cite{dalda08}. The simulations explain
the sea-serpent structure and dynamics of the penumbral field as a
consequence of solar magnetoconvection in a highly inclined, strong
magnetic field, which forms filamentary structures and has properties
of traveling convective wave.

\begin{figure}
\begin{center}
\includegraphics[scale=1.1]{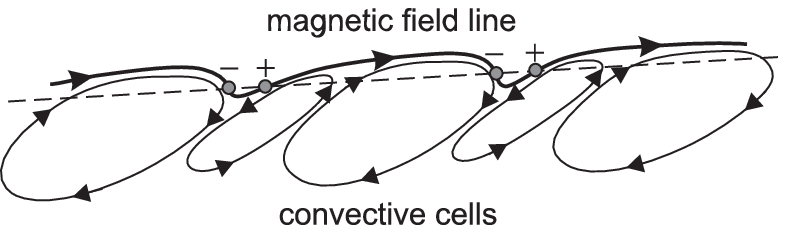}
\end{center}
\caption{Schematic illustration of the sea-serpent model of sunspot penumbra and the moving bipolar magnetic patches. The inclined elliptical curves illustrate streamlines of magnetoconvection in highly inclined magnetic field. The thick solid curve illustrates the sea-serpent behavior of magnetic field lines. Dashed line indicates the solar surface, where the magnetic field is observed.
\label{scheme}}
\end{figure}

The physical picture schematically illustrated in Figure~\ref{scheme}
is the following.  Convective cells in sunspot penumbrae are deformed
under the action of the inclined magnetic field, forming filamentary
structures and producing high-speed Evershed flows
\citep{kiti09a}. The magnetic field lines are stretched by the
downward flows and dragged under the surface. The points where the
magnetic field lines cross the solar surface are observed as magnetic
patches of positive and negative polarities. Note that the negative patch
is closer to the umbra, in agreement with the observations. The
convective cells move in the direction of the magnetic field inclination
because of the traveling convective wave behavior. Therefore, the
bipolar magnetic patches also move in the same direction.

Thus, the numerical simulations connect the sea-serpent structure of
the moving bipolar magnetic pathes observed in the penumbra with the
process of overturning magnetoconvection, traveling convective waves,
and the Evershed flow.

\acknowledgments
This work was partly supported by NASA, the Center for Turbulence Research
(Stanford University), NORDITA, AlbaNova Univ. Center (Stockholm), and
the Spanish Ministerio de Ciencia e Innovaci\'on through project
AYA2009-14105-C06-06 and by Junta de Andaluc\'{\i}a through
project P07-TEP-2687.


\begin{thebibliography}{99}
\bibitem[Beck(2008)]{beck08} Beck, C.\ 2008, \aap, 480, 825.

\bibitem[Bellot Rubio(2009)]{bellot09}
Bellot Rubio, L.R. 2009, in Magnetic Coupling between the Interior and
the Atmosphere of the Sun, ed. S.S. Hasan \& R.J. Rutten, Astrophysics
and Space Science Proceedings, (Berlin: Springer), in press, (arXiv:0903.3619).

\bibitem[Bellot Rubio et al.(2003)]{bellot03}
Bellot Rubio, L. R., Balthasar H., Collados, M. \& Schlichenmaier, R. 2003,
A\&A, 403, L47.


\bibitem[Bellot Rubio et al.(2004)]{bellot04}
Bellot Rubio, L.~R., Balthasar, H., \& Collados, M.\ 2004, \aap, 427, 319.

\bibitem[Borrero et al.(2005)]{borrero05}
Borrero, J.~M., Lagg, A., Solanki, S.~K., \& Collados, M.\ 2005, \aap, 436, 333.

\bibitem[Degenhardt \& Wiehr(1991)]{degenhardt91}%
Degenhardt, D. \& Wiehr, E.\ 1991, A\&A, 252, 821.

\bibitem[Evershed(1909)]{evershed1909}%
Evershed, J. 1909, MNRAS, 69, 454.

\bibitem[Harvey \& Harvey(1973)]{harvey73}%
Harvey, K. \& Harvey, J.\ 1973, Sol.~Phys., 28, 61.

\bibitem[Heinemann et al.(2007)]{heinemann2007}
Heinemann, T., Nordlund, {\AA}., Scharmer, G.~B., \& Spruit, H.~C.\ 2007, \apj, 669, 1390.

\bibitem[Hurlburt et al.(2000)]{hurlburt2000}
Hurlburt, N.~E., Matthews, P.~C., \& Rucklidge, A.~M.\ 2000, \solphys, 192, 109.

\bibitem[Ichimoto et al.(2007)]{ichimoto07}
Ichimoto, K., Shine, R.~A., Lites, B., Kubo, M., Shimizu, T., et al.\ 2007,
PASJ, 59, S593.

\bibitem[Jacoutot et al.(2008a)]{jacoutot08a}
Jacoutot, L., Kosovichev, A.~G., Wray, A.~A. \& Mansour, N.~N.\ 2008a, ApJ, 682, 1386.

\bibitem[Jacoutot et al. (2008b)]{jacoutot08b}
Jacoutot, L., Kosovichev, A.~G., Wray, A.~A. \& Mansour, N.~N.\ 2008b, ApJL, 684, L51.

\bibitem[Kitiashvili et al.(2009a)]{kiti09a}
Kitiashvili, I.~N., Kosovichev, A.~G., Wray, A.~A. \& Mansour, N.~N. 2009a, ApJL, 700, 178.

\bibitem[Kitiashvili et al.(2009b)]{kiti09b}
Kitiashvili, I.~N., Jacoutot, L., Kosovichev, A.~G., Wray, A.~A. \&
Mansour, N.~N. 2009b, Proc. of Int. Conf. Stelar Pulsation: challenges
for theory and observation. AIP Conf. Proc., 1170, 569.

\bibitem[Kitiashvili et al.(2009c)]{kiti09c}
Kitiashvili, I.~N., Kosovichev, A.~G., Wray, A.~A. \& Mansour, N.~N.\ 2009c,
Proc. of ParCFD Conf., NASA Ames, 424.

\bibitem[Kosugi et al.(2007)]{kosugi07}
Kosugi, T., Matsuzaki, K., Sakao, T., Shimizu, T., Sone, Y., et al.\
2007, Sol.~Phys., 243, 3.

\bibitem[Kubo et al.(2008)]{kubo08} Kubo, M., et al.\ 2008,
\apj, 681, 1677

\bibitem[Lites et al.(1993)]{lites93} Lites, B.~W., Elmore,
D.~F., Seagraves, P., \& Skumanich, A.~P.\ 1993, \apj, 418, 928


\bibitem[Rempel et al.(2009a)]{rempel2009}
Rempel, M., Sch{\"u}ssler, M., \& Kn{\"o}lker, M.\ 2009, \apj, 691, 640.

\bibitem[Rempel et al.(2009b)]{rempel_sci09} Rempel, M.,
Sch{\"u}ssler, M., Cameron, R.~H., \& Kn{\"o}lker, M.\ 2009, Science, 325, 171.

\bibitem[Ripoli \& Wray(2003)]{ripoli03}%
Ripoli, J.-F. \& Wray, A.~A.\ 2003, in Annual Research Briefs-2003,
Center for Turbulence Research, Stanford University, 3.

\bibitem[Ravindra (2006)]{ravindra06}
Ravindra B.\ 2006, Sol.~Phys., 237, 297.

\bibitem[Sainz Dalda \& Bellot Rubio(2008)]{dalda08}%
Sainz Dalda, A. \& Bellot Rubio, L.~R.\ 2008, A\&A, 481, L21.

\bibitem[Sainz Dalda \& Mart\'{\i}nez Pillet(2005)]{dalda05}%
Sainz Dalda, A. \& Mart\'inez Pillet, V.\ 2005, ApJ, 632, 1176.

\bibitem[S\'anchez Almeida (2005)]{sanchez05}
S\'anchez Almeida, J. 2005, ApJ, 622, 1292.

\bibitem[S{\'a}nchez Almeida \& Ichimoto(2009)]{sanchez09}
S{\'a}nchez Almeida, J., \& Ichimoto, K.\ 2009, \aap, 508, 963.

\bibitem[Scharmer et al.(2008)]{scharmer2008}
Scharmer, G.~B., Nordlund, {\AA}., \& Heinemann, T.\ 2008, \apjl, 677, L149.

\bibitem[Scherrer et al.(1995)]{scherrer1995}
Scherrer, P.~H., et al.\ 1995, \solphys, 162, 129.

\bibitem[Schlichenmaier(2009)]{schliche09} Schlichenmaier, R.\
2009, Space Science Reviews, 144, 213

\bibitem[Schmidt et al.(1992)]{schmidt92}
Schmidt, W., Hofmann, A., Balthasar, H., Tarbell, T.~D. \& Frank, Z.~A.\ 1992, A\&A, 264, L27.

\bibitem[Stanchfield et al.(1997)]{stanchfield97} Stanchfield,
D.~C.~H., II, Thomas, J.~H., \& Lites, B.~W.\ 1997, \apj, 477, 485.


\bibitem[Sheeley(1969)]{sheeley1969}
Sheeley, Jr.\ 1969, Solar Physics, 9, 347.

\bibitem[Shine et al.(1994)]{shine1994}
Shine, R.~A., Title, A.~M., Tarbell, T.~D., Smith, K., Frank, Z.~A.,
\& Scharmer, G.\ 1994, \apj, 430, 413.

\bibitem[Stein \& Nordlund(2001)]{stein2001}
Stein, R.~F. \& Nordlund, A.\ 2001, ApJ, 546, 585.

\bibitem[Severnyi(1965)]{severnyi65}
Severnyi, A.~B.\ 1965, Soviet Astronomy, 9, 171.

\bibitem[Title et al.(1993)]{title93}
Title, A.~M., Frank, Z.~A., Shine, R.~A., Tarbell, T.~D., Topka,
K.~P., Scharmer, G. \& Schmidt, W.\ 1993, ApJ, 403, 780.

\bibitem[Tsuneta et al.(2008)]{tsuneta2008}
Tsuneta, S., Ichimoto, K., Katsukawa, Y., Nagata, S., Otsubo, M., et
al. 2008, \solphys, 249, 167.


\bibitem[Westendorp Plaza et al.(1997)]{westend97}
Westendorp Plaza, C., del Toro Iniesta, J.~C., Ruiz Cobo, B., Martinez Pillet, V.,
Lites, B.~W., \& Skumanich, A.\ 1997, \nat, 389, 47.


\end{thebibliography}
\end{document}